\documentclass[conference]{IEEEtran}
\IEEEoverridecommandlockouts

\usepackage{cite}
\usepackage{amsmath,amssymb,amsfonts}
\usepackage[ruled,vlined]{algorithm2e}
\usepackage{algorithmic}
\usepackage{graphicx}
\usepackage{amssymb}
\usepackage{subcaption}
\usepackage{pifont}
\usepackage{multirow}
\usepackage{booktabs}
\usepackage{makecell}
\usepackage{textcomp}
\usepackage{enumerate}
\usepackage{xcolor}
\usepackage{xspace}
\usepackage{enumitem}
\usepackage{tikz}
\usepackage{wrapfig,lipsum,booktabs}
\usepackage[hidelinks]{hyperref}

\def\BibTeX{{\rm B\kern-.05em{\sc i\kern-.025em b}\kern-.08em
    T\kern-.1667em\lower.7ex\hbox{E}\kern-.125emX}}

\DeclareMathOperator*{\argmax}{arg\,max}
\DeclareMathOperator*{\argmin}{arg\,min}
    
\newcommand{\NAME}{$\mathtt{AIAShield}$\xspace}    

\newcommand\encircle[1]{\tikz[baseline=(X.base)] 
    \node (X) [draw, shape=circle, inner sep=0pt, fill=black, text=white] {\strut #1};}

\begin{document}

\title{Defense against ML-based Power Side-channel Attacks on DNN Accelerators with Adversarial Attacks}

\author{
    \IEEEauthorblockN{Xiaobei Yan\IEEEauthorrefmark{1},  Chip Hong Chang\IEEEauthorrefmark{2}, Tianwei Zhang\IEEEauthorrefmark{1}}
    \IEEEauthorblockA{\IEEEauthorrefmark{1} School of Computer Science and Engineering, Nanyang Technological University, Singapore}
    \IEEEauthorblockA{\IEEEauthorrefmark{2} School of Electrical and Electronic Engineering, Nanyang Technological University, Singapore}
    \IEEEauthorblockA{xiaobei002@e.ntu.edu.sg, \{ECHChang, tianwei.zhang\}@ntu.edu.sg}
}
\pagestyle{plain}
\maketitle

\begin{abstract}
Artificial Intelligence (AI) hardware accelerators have been widely adopted to enhance the efficiency of deep learning applications. However, they also raise security concerns regarding their vulnerability to power side-channel attacks (SCA). In these attacks, the adversary exploits unintended communication channels to infer sensitive information processed by the accelerator, posing significant privacy and copyright risks to the models. Advanced machine learning algorithms are further employed to facilitate the side-channel analysis and exacerbate the privacy issue of AI accelerators. Traditional defense strategies naively inject execution noise to the runtime of AI models, which inevitably introduce large overheads. 

In this paper, we present \NAME, a novel defense methodology to safeguard FPGA-based AI accelerators and mitigate model extraction threats via power-based SCAs. The key insight of \NAME is to leverage the prominent adversarial attack technique from the machine learning community to craft delicate noise, which can significantly obfuscate the adversary's side-channel observation while incurring minimal overhead to the execution of the protected model. At the hardware level, we design a new module based on ring oscillators to achieve fine-grained noise generation. At the algorithm level, we repurpose Neural Architecture Search to worsen the adversary's extraction results. Extensive experiments on the Nvidia Deep Learning Accelerator (NVDLA) demonstrate that \NAME outperforms existing solutions with excellent transferability.

\end{abstract}

\section{Introduction}
\label{sec:intro}
With the rapid advancement of Artificial Intelligence (AI), hardware accelerators have emerged as a pivotal component in optimizing the execution of complex AI workloads. These specialized hardware architectures unlock the unprecedented computational capabilities, enabling the deployment of AI-powered systems across various domains, including autonomous driving\cite{hao2019nais}, medical analysis\cite{gulfidan2021artificial}, natural language processing\cite{khan2021npe}, etc. However, this surge in the adoption of AI accelerators gives rise to a new set of security challenges. One prominent threat is power side-channel attacks (SCAs), which capitalize on unintended information leaks, i.e., power consumption patterns \cite{wei2018know}, to infer sensitive data processed by the accelerator. Past works have demonstrated that SCAs can be used to facilitate model extraction attacks \cite{yan2023mercury,tian2021remote,li2021power,zhang2021stealing,yoshida2019model}, enabling an external adversary to precisely recover the model attributes (e.g., architecture, hyperparameters, parameters). This significantly jeopardizes the privacy and intellectual property of AI models running on the accelerators.



Various techniques have been proposed to analyze power side-channel traces and recover the sensitive information. Conventional solutions include Differential Power Analysis (DPA) \cite{kocher1999differential} and Correlation Power Analysis (CPA). Although they are proficient at attacking cryptographic algorithms, they lack the adaptability to compromising commercial off-the-shelf AI accelerators due to the distinct designs and execution characteristics. Recently researchers exploit machine learning (ML) algorithms to process side-channel sequences, and extract useful information with high accuracy, generalization, and automation \cite{hettwer2020applications}. Such ML-based profiled SCAs are particularly powerful in attacking AI accelerators which have intricate architectures and noisy execution traces. Recent studies have shown the feasibility of model extraction attacks from power side channels using advanced ML techniques \cite{yan2023mercury}.

It is necessary to develop effective and efficient countermeasures against power side-channel attacks. Prior works have introduced different solutions to protect cryptographic accelerators, which can be classified into two categories. (1) \textit{Masking} \cite{9603186,6296466}. This strategy detaches the power consumption from the actual secret data by randomizing the sensitive variables during its processing. (2) \textit{Hiding} \cite{morid2023shield,krautter2019active,zhang2022darpt,chong2021dual}. This strategy aims to equalize the power consumption throughout the execution, making it difficult to find exploitable information in the leakage. It can be realized via two techniques: clock misalignment (hiding in the time dimension) \cite{morid2023shield,krautter2019active,chong2021dual} and noise compensation (hiding in the amplitude dimension) \cite{zhang2022darpt,chong2021dual}. These solutions are further extended to protect AI accelerators and prevent model extraction from power side-channel attacks \cite{dubey2020bomanet,dubey2022guarding,maji2022threshold,zhang2022pareto}.

Unfortunately, the above methods are ineffective in defeating ML-based side-channel attacks. In particular, masking is usually achieved by separating internal values into multiple statistically independent shares. Since this is not possible for non-linear functions (e.g., activation functions in AI models), ensuring the correctness of the computation requires a large algorithmic overhead. Besides, ML-based attacks utilize higher statistical moments, rendering the masking countermeasure ineffective. Hiding with clock misalignment adds a random waiting time between the execution of instructions, which is also invalid to ML-based attacks due to the shift-invariance characteristic in  CNN architectures \cite{cagli2017convolutional}.
Hiding with noise compensation lowers the signal-to-noise ratio (SNR) in the leakage, normally via adding noise. However, ML-based attacks can still extract meaningful model information even with a low SNR. Simple noise can be removed by attackers using signal processing with repeated measurements.

Alternatively, we can mitigate power side-channel attacks from the algorithm level. Researchers introduce the \textit{obfuscation} strategy, which obfuscates the model architectures to prevent the attacker from extracting the correct one \cite{zhou2022obfunas,li2021neurobfuscator,luo2022nnrearch}. However, this strategy has several limitations:
(1) For each protected architecture, it needs to identify the corresponding obfuscation target, which is time- and cost-inefficient. 
(2) The defender is required to redesign or even re-train the models. This is infeasible in some scenarios, where the model users have no expertise or privileges for model modifications. 
(3) As these solutions perform obfuscations only at the software level rather than the hardware level, the perturbed side-channel trace is restricted, limiting the defense effectiveness. According to \cite{zhou2022obfunas}, ObfuNAS \cite{zhou2022obfunas} achieves only around 1\% to 3\% accuracy degradation for the attacker's extracted model, but also deteriorates the victim's model accuracy by about 1\%; NeurObfuscator \cite{li2021neurobfuscator} even enhances the accuracy of the extracted models by 2.5\% in some circumstances.

To overcome the above limitations, this paper presents \NAME, a novel hardware-based defense approach to protect AI accelerators against ML-based power SCAs. The key idea of \NAME is to \textit{leverage the adversarial attack technique in machine learning to obfuscate side-channel leakage}. Adversarial attack is a well-studied threat to machine learning models \cite{szegedy2013intriguing}, where the attacker injects carefully-crafted imperceptible perturbations into the input data to mislead the target model. In our context, since the attacker can exploit machine learning models to extract information from the power side-channel trace, the defender can also add such perturbations into the trace to deceive the attacker into recovering wrong information. However, there are several challenges to realize such strategy. 

First, it is difficult to identify the goal of misleading the attacker for perturbation generation. Existing works just obfuscate the model architecture to be significantly different from the target one \cite{luo2022nnrearch,zhang2022pareto}. However, as observed in \cite{zhou2022obfunas}, simply maximizing the architecture difference can still allow the attacker to extract a model with equivalent or even improved performance. In \NAME, we propose two defense goals for the defender to choose. (1) \textit{Model Similarity Reduction}. This aims to increase the model extraction errors with the perturbed side-channel trace. We use the FGSM algorithm \cite{goodfellow2014explaining} to identify the perturbations. (2) \textit{Model Utility Reduction}. This aims to mislead the attacker to extract a model with the worst performance. We leverage Neural Architecture Search (NAS) \cite{zoph2016neural} to identify the least-optimal model architecture, and entice the attacker to obtain such a bad-quality model. The perturbation can be generated with adapted PGD \cite{madry2017towards}.

Second, it is challenging to generate the desired execution noise corresponding to the identified perturbations. 
Common solutions for generating noise on FPGAs rely on Ring Oscillators (ROs). However, the switching activity of ROs can lead to substantial voltage overshoot and undershoot, as depicted in Figure \ref{roe}, thereby complicating precise voltage control and quantization. This voltage transient aspect presents a significant hurdle for the successful implementation of more intricate noise patterns using ROs. To address this challenge, we design a fine-tuned FPGA module with novel hardware and software designs, which performs sophisticated calibration to achieve fine-grained noise generation. 

We implement \NAME on Nvidia Deep Learning Accelerator (NVDLA), the mainstream open-source configurable architecture. We perform extensive experiments to validate the effectiveness and superiority of \NAME in several aspects:

\begin{itemize}[leftmargin=*]

\item \textbf{Minimal Area Increase:} \NAME incurs the smallest chip area increase on the FPGA compared to other existing defense strategies \cite{yao2021programmable,dubey2020bomanet,dubey2022guarding,zhang2022darpt}.

\item \textbf{Effective Extraction Accuracy Degradation:} \NAME can significantly reduce the attacker's extraction accuracy compared to the common random noise injection. 

\item \textbf{Effective Model Accuracy Degradation:} \NAME can also force the attacker to obtain a bad-quality model, rendering it less functional. 

\item \textbf{High Transferability and Robustness: } \NAME shows remarkable transferability across a wide range of attack models. It also exhibits robustness when the noise generator is placed at different locations on the FPGA.

\end{itemize}

The remainder of this paper is structured as follows: Section \ref{sec:Background} provides the preliminaries. Section \ref{sec:ProblemFormulation} discusses the problem statement. Section \ref{sec:details} elaborates the detailed design of \NAME. Section \ref{sec:evaluations} presents the evaluation results. Section \ref{sec:relatedworks} gives the related works. Section \ref{sec:discussion} presents some discussions and Section \ref{sec:conclusion} concludes the paper.

\section{Background}
\label{sec:Background}

\subsection{Nvidia Deep Learning Accelerator (NVDLA)}
NVDLA, an open-source configurable architecture developed by Nvidia, is designed to accelerate deep learning inference tasks. It can perform convolution, activation, pooling, and normalization operations during model inference. NVDLA can be customized to function as either a large or small implementation, with variations in core dimensions and specific engine implementations (such as Rubik and DMA engines) \cite{nv_thesis}.

\begin{figure}[t]
	\centering
	\includegraphics[scale=0.33,trim=0 0 0 0]{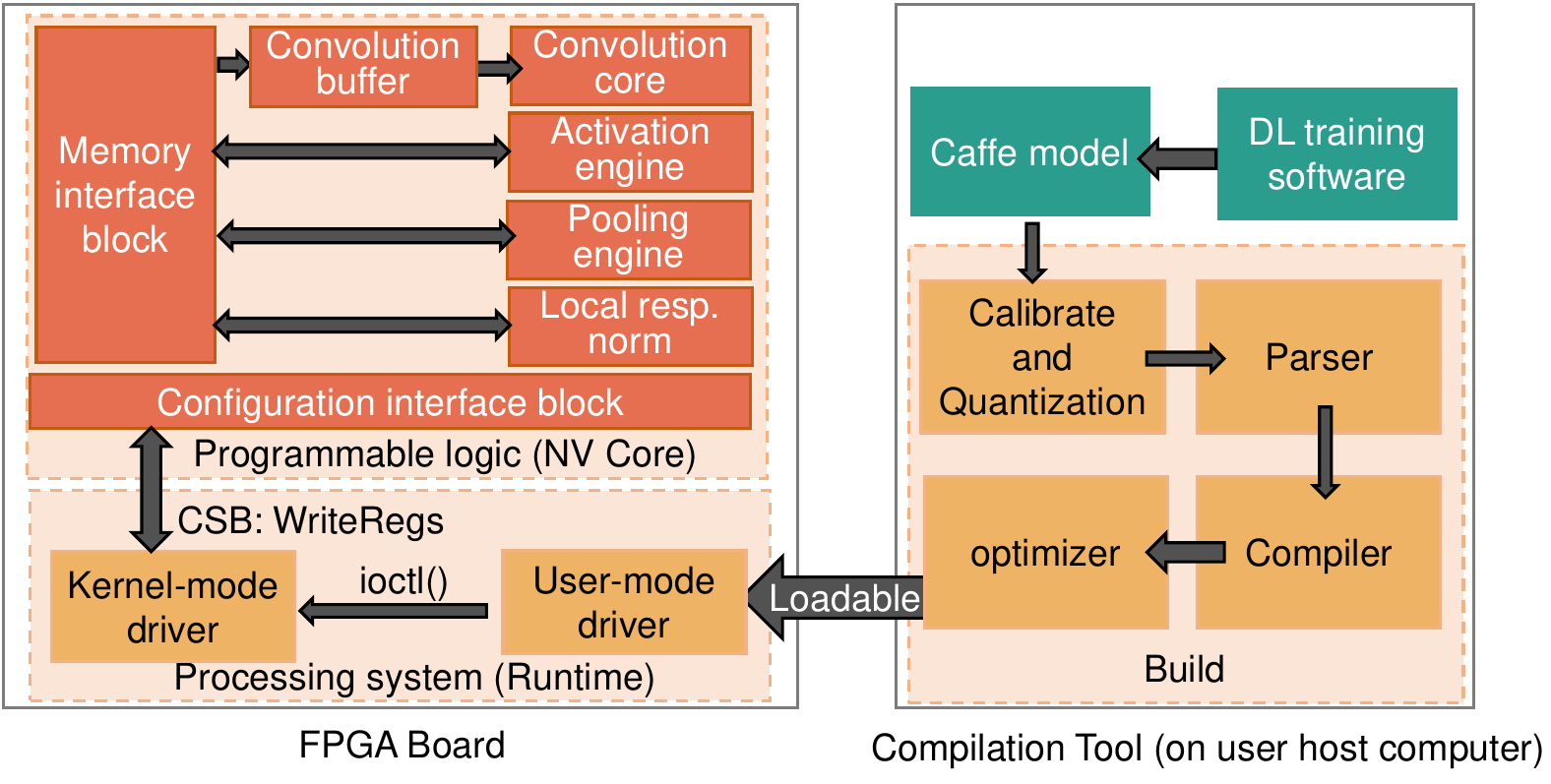}
    \vspace{-15pt}
	\caption{Architecture overview of NVDLA}
	\vspace{-15pt}
	\label{nv}
\end{figure}

The architecture overview of NVDLA is illustrated in Figure \ref{nv}. It can be categorized into two main components: hardware design and software design. The hardware design consists of a sequence of pipeline stages equipped with diverse types of engines, ensuring the efficient operation of FPGA boards. On the other hand, the software design acts as an intermediary connecting users and hardware components. It is responsible for constructing and loading the Deep Neural Network (DNN) model onto the FPGA for execution.

The software design is further divided into two components: (1) Compilation Tools: they take the pre-compiled model from Caffe and generate a network of hardware layers compatible with NVDLA. This network, termed "loadable," is then calibrated using TensorRT calibration.
(2) Runtime Environment: this component handles the calibrated loadable and directly executes it within the NVDLA environment.

\subsection{Machine Learning-based Side-channel Attacks}
\label{sec:background-sca}

The advance of machine learning (ML) algorithms has significantly propelled the field of side-channel analysis \cite{standaert2009compare}. Attackers can utilize ML models to automate and optimize the extraction of sensitive information from complex side-channel traces. Typically, such attacks unfold in two phases.

\textbf{Profiling Phase:} The attacker executes the target application on either the same or a similar device and learns the physical leakages. If the attacker possesses an input secret set denoted as $ \mathbf{s}=\{s_1,...,s_n\} $, he can collect $N$ side-channel traces $ \mathbf{{T_{i,n}}} $ corresponding to each input $ s_i $. Using such data, he constructs an ML model $ \mathit{f}: \mathbf{T_{i,n}}\mapsto s_i $, effectively establishing a correlation between the side-channel traces and input secrets.

\textbf{Exploitation Phase:} Armed with the developed ML model $\mathit{f}$, the attacker is ready to exploit the acquired knowledge at runtime. By utilizing an additional set of \textit{q} traces $ {T}'_1,...,{T}'_q $ captured from the targeted device, the attacker can infer the secret $ s^{'}_i $ by evaluating $ s^{'}_i=\mathit{f}({T}'_i) $.

ML-based side-channel attacks offer several advantages. Firstly, they can retrieve information even when discernible patterns in the power trace are absent. This is particularly relevant in cases where computations on FPGAs occur in parallel, making manual pattern analysis ineffective. Secondly, ML models can seamlessly integrate all the information contained within a single power trace, as they excel at handling high-dimensional data. In contrast, traditional methods necessitate the identification of specific points of interest (POI) to narrow down the information for the attack. Thirdly, ML models exhibit greater robustness against noise present in the side-channel trace compared to conventional statistical techniques. This robustness enhances their efficacy in real-world scenarios.

Yan et al. \cite{yan2023mercury} realize such attack against NVDLA. The attacker uses a Time-to-Digital Converter (TDC) to collect power side-channel traces, and trains a sequence-to-sequence model to extract the victim's model architectures. The attacker's model incorporates three CNN layers responsible for extracting features from the input data, alongside two RNN layers comprising 128 dimensions each, which facilitate the information propagation. To bolster its ability to retain long-term memory, the RNN component employs the bidirectional gated recurrent unit (BiGRU) architecture. As a result of this arrangement, the RNN layer generates a probability distribution corresponding to each input instance, which is then transmitted to the CTC decoder for further processing.

\subsection{Adversarial Attack against Machine Learning Models}
\label{sec:adversarial-attack-background}
Adversarial attacks \cite{szegedy2013intriguing} refer to the intentional manipulations of input data with the objective of misleading machine learning models to produce incorrect outputs. They have become one of the most severe threats to machine learning applications and systems. Researchers have proposed various attack techniques targeting different domains, including images \cite{szegedy2013intriguing}, language texts \cite{morris2020textattack}, audios \cite{carlini2018audio}, etc. 

There are two primary categories of attacks based on the attacker's goals. (1) \textit{Untargeted attacks}. They are designed to cause misclassification of the victim model without any specific objective in mind. By crafting special imperceptible perturbations, the malicious input could make the model predict any incorrect class label with very high confidence. (2) \textit{Targeted attacks}. The attacker generates adversarial perturbations that lead the victim model to misclassify the input into a specific predefined class label chosen by him. \NAME leverages both attack techniques to achieve two defense goals. Specifically, with the untargeted attack technique, we achieve Model Similarity Reduction, which could increase the attacker's model extraction error rate. With the target attack technique, we achieve Model Utility Reduction, which entices the attacker to extract a model with the worst performance.

\subsection{Neural Architecture Search (NAS)}
The success of deep learning heavily relies on finding the optimal model architectures tailored to specific tasks. However, manual architecture design and tuning can be laborious and time-consuming. NAS, a subfield of automated machine learning (AutoML), offers a breakthrough by automating the process of architecture design, enabling the discovery of high-quality models with minimal human intervention. \cite{zoph2016neural}

NAS defines a search space as the scope of neural networks in consideration, from which it
finds the best architecture with different types of search algorithms, such as evolutionary algorithms\cite{miller1989designing}, reinforcement learning\cite{zoph2016neural}, or gradient-based optimization\cite{miikkulainen2019evolving}. Instead of relying on human experience, NAS utilizes computational resources to efficiently navigate through the design space, optimizing for performance metrics like accuracy, model size, or computational efficiency. In \NAME, we refine the NAS algorithm to search for the \textit{worst-performing} models instead of \textit{best-performing} ones, set as the target for the attacker to extract. 

\section{Problem Formulation}
\label{sec:ProblemFormulation}

\subsection{Threat Model}

Following the common threat model in prior works \cite{tian2021remote,moini2021remote,luo2021deepstrike}, we consider the multi-tenant cloud-FPGA services, e.g., AWS F1 instance \cite{amazon} and Microsoft Azure \cite{microsoft2018}. The cloud provider allocates multiple users' implementations on the same FPGA board to enhance the resource utilization and elasticity. In such environment, different users' applications are logically isolated and running concurrently, which is guaranteed by the FPGA virtualization technology \cite{10.1145/3373376.3378491}. Meanwhile, they also share specific hardware resources, such as Power Distribution Network (PDN), hence enabling the side-channel leakage. 

The victim developer implements an AI accelerator (i.e., NVDLA) on one cloud FPGA instance, and loads the protected AI model for inference. A malicious user performs the co-location attack to launch his circuit application on the same FPGA chip. 

With such setup, the attacker can conduct remote side-channel attacks introduced in Section \ref{sec:background-sca} to extract the architecture details of the victim model. Specifically, he can implement his circuit to monitor the power consumption activities on the board, which are heavily affected by the runtime execution of the victim model. By performing the profiling and exploitation phases, the attacker can build an ML model $f$ to automatically recover victim's model architecture from the power side-channel trace $T$: $\mathrm{M} = f(T)$. Past works have validated the feasibility and severity of such attacks \cite{yan2023mercury}.

\subsection{Defense Goals and Requirements}
To prevent the extraction of the valuable model, the defender could leverage adversarial attack algorithms to generate noise $\Delta t$ (i.e., voltage fluctuation) and inject it to the execution of the model inference to obfuscate the side-channel traces. By carefully adjusting the scale of the noise and injection moments, the defender can achieve different effects. Inspired by the untargeted and targeted attacks discussed in Section \ref{sec:adversarial-attack-background}, we propose two defense goals: 

\textbf{1. Model Similarity Reduction}: the defender aims to make the attacker's extracted model as distinct from the correct one as possible. This goal can be formulated as follows:
\begin{equation}
\argmax_{\Delta t} L(f(T+\Delta t), \mathrm{M})
\end{equation}
where $L$ represents a distance function that measures the model difference, \textit{f} represents the attacker's extraction model, and M is victim's model deployed on the accelerator.

\textbf{2. Model Utility Reduction}: the defender aims to make the attacker's extracted model have as low accuracy as possible. This goal can be formulated as follows:
\begin{equation}
\argmin_{\Delta t} \texttt{Acc}(f(T+\Delta t))
\end{equation}
where $\texttt{ACC}$ measures the testing accuracy of a model.

A practical defense must satisfy the following requirements:

\begin{itemize}[leftmargin=*]

\item \textit{Effectiveness}. The injected noise can effectively obfuscate the attacker's side-channel observation such that the extracted model has significant difference from the original one, or very low accuracy. 

\item \textit{Computation efficiency}. The defense solution incurs minimal impact on the computation of the FPGA board. On the one hand, the new hardware component should introduce very small area and power consumption on the board. On the other hand, the injected noise should hardly affect the latency or accuracy of the victim model. 

\item \textit{Implementation-friendly}. The defender only needs to implement a small hardware module for defense. He does not need to modify or customize the target model. This is important in lots of scenarios where the defender does not have the privilege or expertise to touch the model. 

\item \textit{Generalization}. For a given task, the implementation is general and can be applied to arbitrary models. 

\item \textit{Black-box defense}. The defender does not know the detailed ML model employed by the attacker. Instead, he can adopt a local surrogate model to craft the perturbation noise. This surrogate model can be constructed from a pre-trained model on the Internet, or by training from scratch. Due to the transferability property of adversarial attacks \cite{demontis2019adversarial}, the generated noise can also defeat attacker's actual model. A comprehensive evaluation of such transferability can be found in Section \ref{transferabilitysection}.

\end{itemize}

\section{\NAME}
\label{sec:details}

\subsection{Overview}
\begin{figure}[t]
	\centering
	\includegraphics[width=\linewidth]{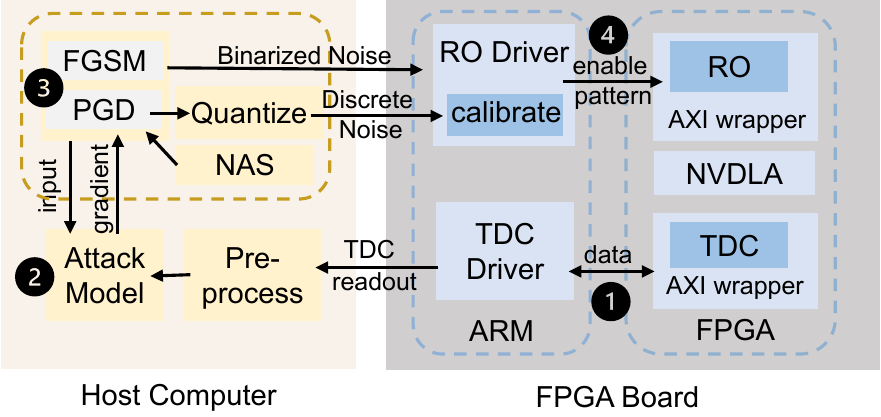}
 
	\caption{\NAME framework and its workflow.}
    \vspace{-15pt}
	\label{fw}
	
\end{figure}

The overview of \NAME is illustrated in Figure \ref{fw}. Its key idea is to leverage adversarial attack techniques to craft noise, which can obfuscate the attacker's side-channel observations. To achieve this goal, we make innovations at both the hardware and algorithm levels. 

At the hardware level, we implement Ring Oscillators (ROs), which receive the required scale of noise from the host computer, and incur the corresponding voltage fluctuations into the side-channel power trace. The calculated noise with adversarial attack techniques is very delicate such that even a small shift could make it ineffective. Hence, we need ROs to generate very fine-grained noise on demand, which is challenging. To this end, we introduce a novel fine-grained Noise Generation Module with a software calibration scheme (Section \ref{sec:noise-generation}). Besides, we implement a Time-to-Digital Converter (TDC), which is used by the defender in the offline phase, to collect the model execution trace for noise calculation (Section \ref{sec:tdc}). 
Effective communication between the host computer and FPGA is facilitated through the driver programs executed on the ARM processor on the board, which control the FPGA via AXI buses.

At the algorithm level, we introduce new solutions to calculate the execution noise to achieve different defense goals. In particular, for the \textit{Model Similarity Reduction} defense, we adopt the untargeted adversarial attack solution, and employ the Fast Gradient Sign Method (FGSM) \cite{goodfellow2014explaining} algorithm to generate adversarial noise. By mapping the noise to the corresponding enable count for ROs, we can increase the attacker's model extraction error rate. Since FGSM produces noise with only two values (i.e., $\pm\epsilon$), the mapping to RO enable and disable states is straightforward (Section \ref{untargeted defense}).

For \textit{Model Utility Reduction}, we turn to the targeted adversarial attacks. We first utilize a modified Neural Architecture Search (NAS) algorithm to search for the model architecture with the worst performance on the given task. Subsequently, we employ our adapted Projected Gradient Descent (PGD) algorithm \cite{madry2017towards} to generate the desired noise, which deceives the attacker into recovering the bad-quality model from the perturbed trace. Unlike FGSM which yields binary noise values, our tailored PGD generates a continuous spectrum of noise values. To harmonize this diverse noise range, we quantize the calculated noise into discrete values. We then use the fine-grained calibration scheme to translate the discrete noise into different enabling patterns of the Noise Generation Module, ensuring precise alignment between the noise generation process and side-channel obfuscation (Section \ref{targeted defense}).

\noindent\textbf{Workflow.}
\NAME operates in two phases. In the \textit{offline} phase, the defender trains a surrogate attack model, and calculates the required noise. 
To achieve this, he deploys a TDC and different models on a FPGA board with the same configurations as the actual one, and collects the TDC readouts as the side-channel traces ({\tiny\encircle{\normalsize1}}). Based on the traces and model attributes, he is able to train the attack model ({\tiny\encircle{\normalsize2}}). 
With this surrogate model, the defender can use our algorithms to generate the corresponding noise for different defense goals ({\tiny\encircle{\normalsize3}}).  It is worth noting that the surrogate model can have a different network architecture and hyperparameters from the attacker's actual one, but the generated adversarial noise has strong transferability to defeat different attacks.

In the \textit{online} phase, the defender implements ROs along the protected model, and instructs the ROs to generate the required noise during the model inference execution ({\tiny\encircle{\normalsize4}}).

\subsection{Power Monitor Module}
\label{sec:tdc}
In the offline phase, the defender needs to implement a Power Monitor module to collect the execution trace for building the surrogate model. Following \cite{yan2023mercury}, we employ a Time-to-Digital Converter (TDC) as the power sensor. The TDC operates by capturing the combinational logic delay, leveraging the propagation of a clock signal through a chain of buffers to detect voltage fluctuations. Due to variations in switching activities across different computational operations of the FPGA, discrepancies in voltage drop values may arise, consequently leading to different delay measurements in the TDC. Through this mechanism, we can infer the activities of adjacent circuits using the TDC's readout. 
\begin{figure}[t]
	\centering
	\includegraphics[width=\linewidth]{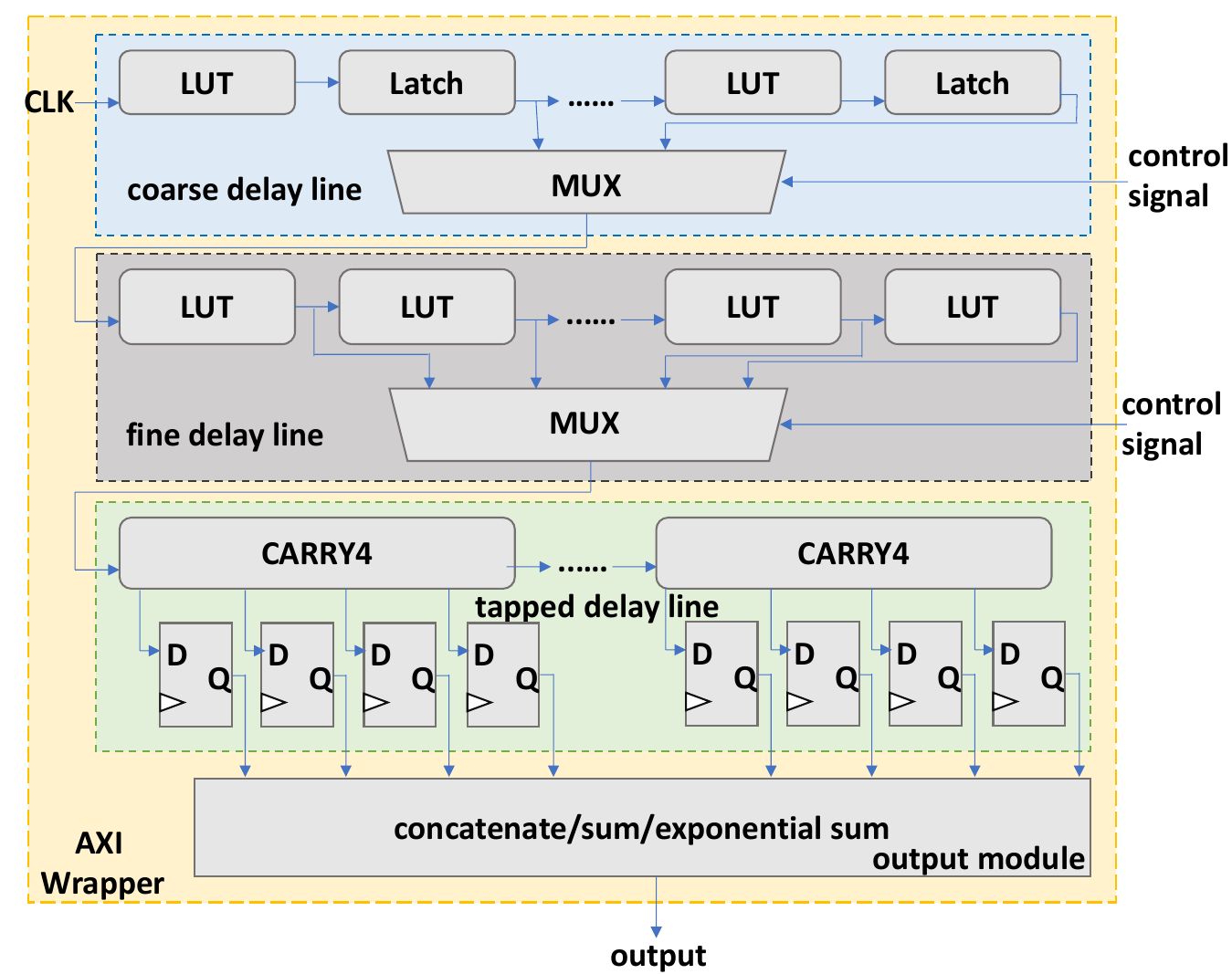}
 
	\caption{Architecture details of TDC.}
    \vspace{-10pt}
	\label{tdc-detail}
	
\end{figure}

Figure \ref{tdc-detail} provides a detailed view of the TDC architecture. In this design, the clock signal enters the TDC and encounters an adjustable coarse delay line and a fine delay line. These components work together to give an initial delay, which is subsequently fed into a tapped delay line. The flexibility of the initial delay is achieved through dynamic configuration facilitated by multiplexers (MUX). By altering the number of logic elements that make up the coarse and fine delay lines during the calibration process through MUXs, the delay duration can be customized. 

The coarse delay line is composed of replicated look-up table (LUT) and latch modules, providing a substantial delay. Conversely, the fine delay line incorporates replicated LUT modules, offering a finer degree of delay control. The tapped delay line employs carry chains and is constructed using \verb!CARRY4! primitives, with their \verb!CO! outputs registered by four dedicated D flip-flops. During each readout, this component tracks the taps reached by the clock signal, providing a raw value. Depending on the configuration specified in the TDC IP settings, this raw output can be concatenated or transformed into a sum or exponential sum.

It is worth highlighting the significance of TDC calibration, particularly the adjustment of its initial delay, prior to conducting output measurements. Our calibration process is implemented in two loops within our TDC driver. It systematically explores all possible combinations of fine and coarse delay line lengths to determine the optimal initial delay value. This ensures that the clock signal remains within the delay line when the state of each D flip-flop is captured by the register.

\subsection{Fine-Grained Noise Generation Module}
\label{sec:noise-generation}

\begin{figure}
  \centering
		\subfloat[An RO unit\label{ro:a}]{
			\includegraphics[scale=0.23]{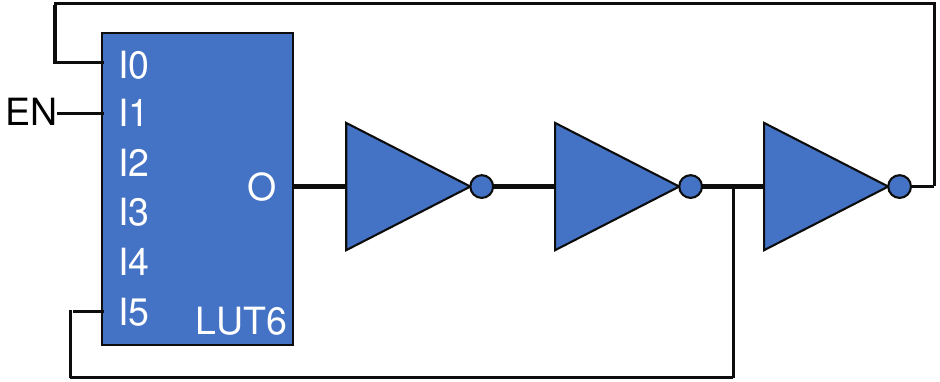}}
		\subfloat[Noise Generator Module\label{ro:b}]{
			\includegraphics[scale=0.27]{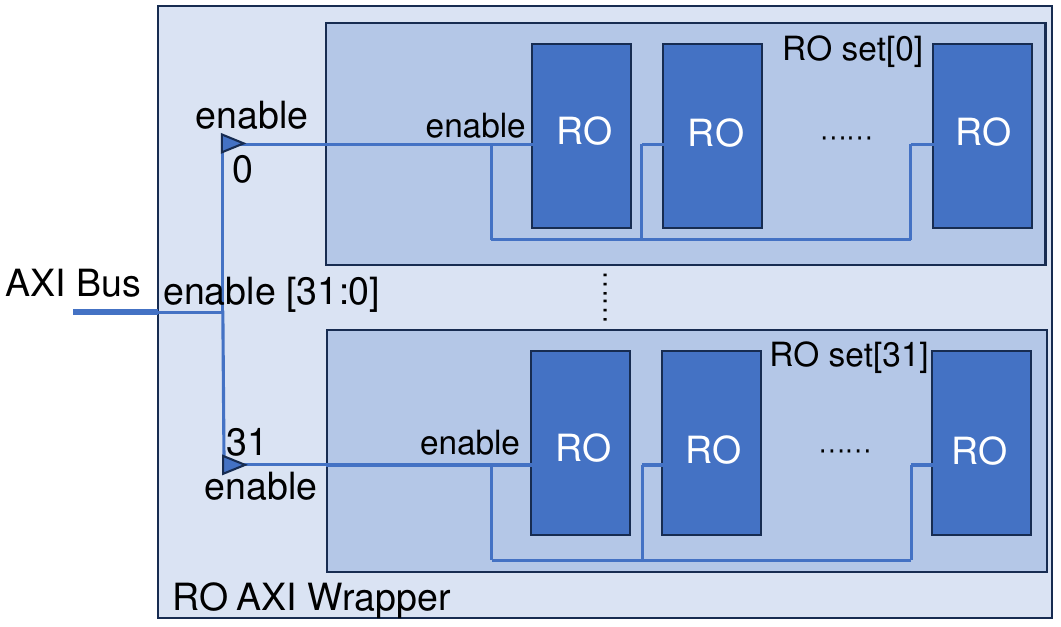}}
\caption{Hardware details of RO and Noise Generation Module}
\vspace{-15pt}
\label{ro}
\end{figure}

\subsubsection{Hardware Design}
In \NAME, we use Ring Oscillators (ROs) to generate noise. An RO consists of an odd number of inverters, with the output of the last inverter fed back to the first one. The prototype of our noise generator is based on a previous work \cite{8740832}. Figure \ref{ro:a} provides the architecture detail, where a 6-input look-up table (LUT) is connected with 3 inverters in series, and an additional feedback path links the last inverter to the LUT. This feedback mechanism is able to prevent glitches. If the enable signal experiences a brief downturn shortly after a falling edge is produced in the additional feedback signal (resulting in a rising edge on the RO output), the output of this LUT is set to a constant 1 only after the additional feedback path changes again.

In our design, a single RO unit requires 4 LUT elements. We assemble 64 RO units to form a set, and the Noise Generation Module comprises 32 such sets, as shown in Figure \ref{ro:b}. To control the enable and disable operations of all sets effectively, we utilize a 32-bit-long AXI bus, with each bit in the bus corresponding to one set of ROs. The AXI bus is accessible by the ARM processor, allowing for read and write operations to its allocated address. This ensures efficient control and coordination of noise generation for the defense. 

\begin{figure}[t]
	\centering
        \includegraphics[scale=0.44,trim=0 0 0 0]{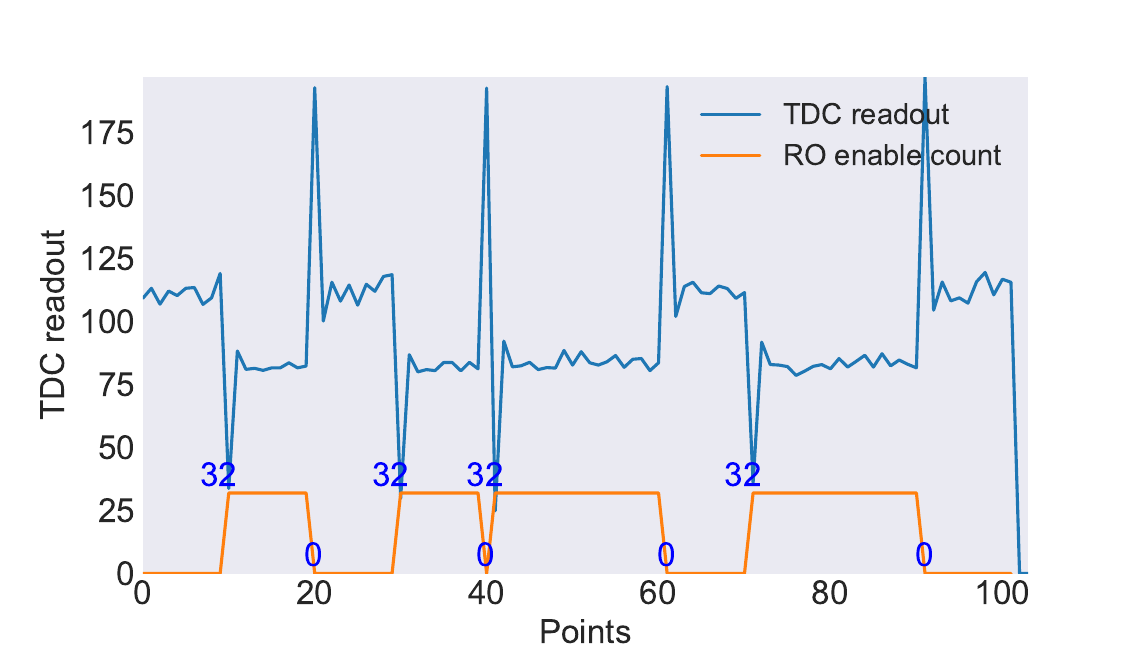}
	\caption{Overshoot and undershoot observed in TDC readout }

    \vspace{-15pt}
	\label{roe}
\end{figure}

\subsubsection{Software Calibration}
A couple of challenges exist to achieve fine-grained noise generation. First, the switching activity of ROs introduces undershoot or overshoot in the TDC readout \cite{10.1145/2435264.2435283}, making it difficult to generate accurate noise. To show this phenomenon, Figure \ref{roe} presents the change of TDC readout (blue line) when the noise generator module is enabled. The orange line represents the settings of enabling (non-zero value) and disabling (zero value) the noise generator module. It is clear that when the noise generator is enabled, the TDC readout drops immediately with a short undershoot. After the RO is disabled, the TDC readout has a short overshoot and then returns to normal. This observation is consistent with the conclusion in \cite{10.1145/2435264.2435283} (Figure 3). Second, for the implementation of noise patterns that involve a sequence of distinct enable counts for ROs, the voltage transient may have a correlation with the preceding RO enable count, thus complicating the faithful realization of the intended noise profile. 

\begin{figure*}[t]
	\centering
        \includegraphics[scale=0.95,trim=0 0 0 0]{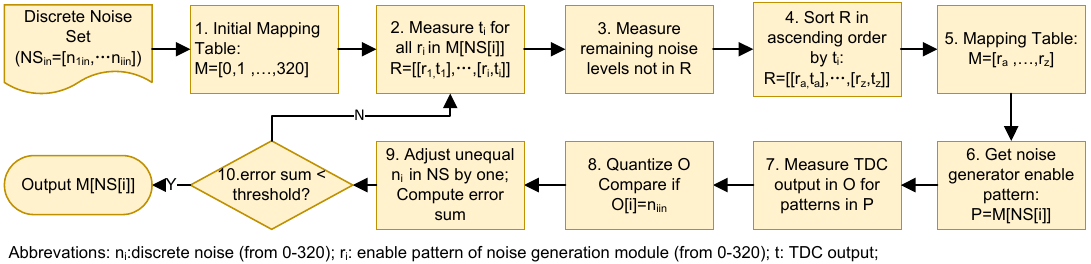}
	\caption{Software calibration in the Noise Generation Module}

    \vspace{-15pt}
	\label{cal}
\end{figure*}

To address these concerns, we design a software calibration and mapping scheme through an iterative approach. For precise measurements, we collect 10 consecutive TDC readouts and calculate their average. We then divide the 321 different enable patterns $r_i$ of the Noise Generation Module. Each pattern corresponds to a specific combination of 32 different RO set enable numbers denoted as $C$ and 10 different enable times denoted as $T$ (ranging from enabling once in 10 TDC measurements to enabling all 10 times). 

The details of the calibration and mapping scheme are shown in Figure \ref{cal}. It consists of the following steps. 
(1) The mapping table $M$ is initialized as [0, 1, ..., 320].
(2) Given that the voltage transients may have a correlation with the previous state of the module, we measure the TDC readout $t_i$ corresponding to the noise generation patterns $r_i$, ensuring that they maintain the same appearance order as the noise set.
(3) Additionally, we measure TDC readouts for noise levels absent from the original set. All [$r_i, t_i$] pairs form the measurement result array $R$.
(4) We sort all [$r_i, t_i$] pairs in $R$ based on $t_i$ in an ascending order.
(5) Based on the sorting results, the mapping table $M=[r_a,...,r_j]$ is constructed, where the \textit{j}-th noise generation pattern $r_j$ corresponds to the discrete noise level $n_j$.
(6) To apply the mapping table to the discrete noise set \textit{NS}, we use $M[NS[i]]$ to obtain the noise generation pattern for each noise data point.
(7) We measure these noise generation patterns and quantize the measured readings into 321 levels.
(8) We then compare these measured results with the input noise set $NS_{in}$. 
(9) If inconsistencies arise, we adjust the unequal ones in \textit{NS} by one.
(10) To validate the calibration effectiveness, we introduce the "error sum" as an evaluation parameter, which sums up all inconsistency values. It is defined as: $Error\ Sum=\sum_{i=1}^{n} \left | n_{i_{in}}-O[i] \right |$, where $n_{i_{in}}$ is the input noise level in \textit{NS}, and $O[i]$ is the actual measured noise level in each iteration. This process continues until the error sum falls below an established threshold, ensuring the accuracy of the calibration process. Ultimately, the program outputs the calibrated noise set as ${M[NS]}$.

\subsection{Model Similarity Reduction Defense}
\label{untargeted defense}
The objective of this defense is to maximize the attacker's extraction error rate from the side-channel leakage. Given the protected model $\mathrm{M}$, we execute its inference by $n$ times and collect the corresponding power traces ${T_{1},..., T_{n}}$. For each trace $T_i$, we employ FGSM over the surrogate model $f$ to generate the corresponding perturbation $\Delta T_i$, which causes the lowest prediction accuracy of $f$. By combining all perturbations through the computation of $sign(\sum_{i=1}^n\Delta T_{i})$, we obtain a perturbation vector consisting of $\pm1$ values.

For practical implementation, we generate a loadable file compatible with the Noise Generation Module. Since enabling the ROs induces a voltage drop and consequently reduces the TDC readout (Figure \ref{roe}), we map the perturbations as follows: a perturbation value of 1 is linked to level 0 in the Noise Generator Module, which disables all ROs, while a value of -1 corresponds to the desired RO set enable count. This mapping bridges the gap between noise calculation and RO implementation, allowing us to introduce the calculated noise into the model inference execution as desired.

\subsection{Model Utility Reduction Defense}
\label{targeted defense}
The objective of this defense is to mislead the attacker to extract a model which has the worst performance. To this end, we first use an adapted NAS algorithm to search for a model with the lowest prediction accuracy. Then we adopt an adapted PGD algorithm to generate the noise that can make the attacker obtain this searched targeted model. 

\subsubsection{Adapted NAS Algorithm}
Typically, NAS aims to discover the best-performing architecture for a given dataset. In our defense strategy, we want to identify the worst architecture deliberately to deceive the attacker. To achieve this, we can reverse the metric used in NAS to guide the search process to the opposite direction. This can be formulated as follows:
\begin{equation}\label{Eq1}
    \pi^*=\mathop{\arg\max}_{\pi_i \in \Omega}\mathcal{L}(W_{\pi_i};X_{val})
\end{equation}
where $W_{\pi_i}$ is the network parameters associated with the model $\pi_i$,  $\mathbf{\Omega}$ is the search space supported by the accelerator, $X_{val}$ is the validation dataset, $\mathcal{L}(\cdot)$ stands for the loss function, $\pi^{\ast }$ is the target model we expect to find out. 

Researchers have proposed different algorithms to achieve efficient and effective NAS. Without loss of generality, we choose NAS-RL \cite{zoph2016neural}. This solution is based on reinforcement learning, and uses the reward (i.e., accuracy on the validation set) to update the controller RNN. The controller RNN outputs the hyper-parameters of the target model. Validation accuracy is the supervision for training the controller RNN. Therefore, we adjust the reward from $acc_{val}$ to $1-acc_{val}$, to find the worst model in the search space. 

\subsubsection{Adapted Universal PGD Algorithm}
After obtaining the target architecture, generating the corresponding perturbations also exposes unique challenges. Since the same model can lead to distinct side-channel traces with inherent noise, we need to identify the adversarial noise that can be uniformly applied to all the traces. Our initial attempts to directly apply the solution in Section \ref{untargeted defense} (i.e., generating noise for different traces and then averaging them) yield unsatisfactory results for new traces, as the targeted attack noise is more sensitive and less general than the untargeted one. 

Inspired by Universal Adversarial Perturbation \cite{moosavi2017universal}, we develop an adapted universal PGD algorithm, to improve the effectiveness of our defense strategy (Algorithm \ref{algo1}). The key aspect of our approach is to generate a universal perturbation by accumulating perturbations from all the side-channel traces. This accumulated perturbation is then used as the common perturbation for all the traces sharing the same label.

\begin{algorithm}[t]
	\caption{Adapted Universal PGD Algorithm.}
	\label{algo1}
	\KwIn{Attack Model $\mathcal{A}$, Trace Set $\mathcal{T}=\{T_{1},..., T_{i}\}$}
	\KwOut{Adversarial noise $\delta$.}  
	\BlankLine
        Initialize $\delta, \delta^{'}$ to all zeros;

	\ForEach{trace $T_{i}$ in $\mathcal{T}$}{
  
            $T_{i}^{'}\leftarrow{T}_{i}+\delta$;

            ${T}_{i}^{'}, \delta^{'} \leftarrow PGD(\mathcal{A}, \mathcal{T}_{i})$;

            $\delta\leftarrow\delta+\delta^{'}$;

            Clamp $\delta$;
	}
        return $\delta$.
\end{algorithm}

\subsubsection{Quantization}
Unlike the defense strategy with FGSM in Section \ref{untargeted defense}, the noise generated by the adapted universal PGD is continuous in nature. However, our Noise Generation Module can only implement noise with discrete levels. Consequently, we need to quantize the continuous noise into these discrete levels. We introduce two quantization methods: linear quantization and non-linear quantization.

Linear quantization feeds the continuous noise into linear functions. The output of these functions is then rounded to fit within the discrete levels ranging from 0 to 320. Non-linear quantization employs the roundings of non-linear convex mapping functions (e.g., $y=x^2$) to assign lower levels to the noise. The intention behind non-linear quantization is to reduce the power consumption by assigning lower levels to noise values, thereby decreasing the enable count of ROs. However, it is important to acknowledge the trade-off between the power overhead and defense effectiveness. In the following, we mainly present the evaluation results with linear quantization.

\section{Evaluation}
\label{sec:evaluations}

\subsection{Experiment Setup}
\noindent\textbf{Testbed.}
We utilize the Xilinx Zynq-7000 SoC ZC706 board (xc7z045ffg900-2) with NVDLA as our testbed for the implementation and evaluation of \NAME. Notably, our defense approach is independent of the AI accelerator architectures and platforms. Due to hardware limitations, we opt for the small implementation of NVDLA. The ARM processor on the board operates on Ubuntu 16.04 OS, which supports NVDLA along with the driver for the Power Monitor and Noise Generation Module. Hardware design is done using Vivado 2019.1. The NVDLA clock frequency is set to 10MHz, while the TDC operates at 150MHz, and the AXI bus of TDC operates at 10MHz. TDC readouts are sent from the board to the host computer through Ethernet using the \verb!scp! command. To train the surrogate model and generate adversarial noise, we employ Pytorch (version 1.13) and CUDA (version 11.6) running on a server equipped with an Nvidia GeForce RTX 3090 GPU. 

\noindent\textbf{Datasets and Models.}
We consider the image classification tasks over the MNIST and CIFAR-10 datasets. For each dataset, we run the evaluation on 200 randomly generated models. These models have random numbers (within the range of [2, 16]) of network layers. The composition of each layer is also determined through random selection. The selection space includes 12 convolution layers (with the kernel size of 2, 3, 4, 5 and output size of 10, 20, 30), 4 pooling layers (with the kernel size of 2, 3, 4, 5), 5 fully-connected layers (with the output size of 100, 200, 300, 400, 500), 1 ReLU layer, and 1 Softmax layer. The initial pre-training of these models is conducted using Caffe. Subsequently, calibration is performed utilizing TensorRT, followed by compilation through the NVDLA compiler. These models are generated on the host computer and subsequently executed on the FPGA using NVDLA runtime.

\noindent\textbf{Defense Targets and Baselines.}
We focus on ML-based power side-channel attacks against AI accelerators, and representative works are \cite{yan2023mercury,cryptoeprint:2023/368}. As the attack proposed in \cite{cryptoeprint:2023/368} still requires manual processing, and can only extract single layers instead of end-to-end model extraction, we mainly select the attack in \cite{yan2023mercury} as our defense target.

We select three state-of-the-art defense solutions as baselines for comparisons: a random noise-based defense \cite{zhao2018fpga}, a sensor-based active defense \cite{krautter2019active}, and a singular defense proven to resist ML-based SCAs \cite{pothukuchi2021maya}. Notably, for \cite{zhao2018fpga}, the random noise is generated using the \verb!randint! function in the Python random library. For \cite{krautter2019active}, its solution controls the noise generator based on readings from the Power Monitor module to flatten the power curve. As enabling the noise generator lowers the Power Monitor reading, we utilize positive feedback to control the noise generator. For \cite{pothukuchi2021maya}, Gaussian sinusoid noise is added to the signal to hide power information. To establish a fair evaluation, we similarly generate the runtime noise following the Gaussian sinusoid distribution and implement it using our fine-grained Noise Generation Module. The Gaussian sinusoid noise is the addition of a sinusoid and Gaussian noise, and its value at any time \textit{T} is generated by:
\begin{equation}
    Offset + Amp \times sin(\frac{2\pi \times T}{Freq} + Noise(\mu, \sigma))
\end{equation}
where the parameters of \textit{Offset}, \textit{Amp}, \textit{Freq}, $\mu$ and $\sigma$ keep changing. \textit{Noise} refers to Gaussian noise.

\noindent\textbf{Metrics.}
We adopt two evaluation metrics to demonstrate the effectiveness of \NAME. For \textit{Model Similarity Reduction}, as the model architecture is represented as a  variable-length sequence with each element denoting the type of the layer, we embrace the concept of Layer Error Rate (LER), which finds its similarity with the Word Error Rate (WER) metric commonly used in sequence-to-sequence tasks. This metric is calculated as $ LER = L({s}',s)/\left \| s \right \| $, where $ \left \| s \right \| $ is the sequence length of $ s $, and $ L({s}',s) $ is the edit distance (Levenshtein) between the ground-truth model layer sequence $ s $ and predicted model layer sequence ${s}'$. A larger LER indicates lower model similarity, thus higher defense effectiveness. For \textit{Model Utility Reduction}, we train each extracted model under the same condition and measure the accuracy of the extracted model. A lower accuracy indicates higher defense effectiveness.

\subsection{Resource Utilization}
In our design, each RO is comprised of only 4 LUTs. With each set containing 64 ROs and a total of 32 such sets, our design efficiently employs only 8,192 LUTs. Factoring in peripheral circuits such as the AXI bus, the aggregated resource utilization for our design stands at 8,251 LUTs and 170 flip-flops (FF). A comprehensive comparison of overhead across various defense solutions from prior works is presented in Table \ref{util}. The ``Area Increased'' column in the table is computed as the ratio of the number of LUTs and flip-flops used by the protective measures to the corresponding counts in the circuits under protection. The ``Utilization'' column computes the proportion of LUT and flip-flop counts relative to the available resources on the FPGA. 
For schemes targeting the AES accelerator, the resource usage is estimated based on the implementation presented in \cite{gravellier2019high}, where 2,640 LUTs and 1,682 FFs are utilized (as the information is not disclosed in these papers). The table demonstrates that among all schemes, \NAME exhibits the least resource consumption in relation to the circuit under protection.
\begin{table}[htbp]
  \centering
  \caption{Overhead comparisons}
  \resizebox{\linewidth}{!}{
    \begin{tabular}{lrrrrll}
    \toprule
    Solution & \multicolumn{1}{l}{LUT} & \multicolumn{1}{l}{FF} & \multicolumn{1}{l}{BRAM} & \multicolumn{1}{l}{Utilization} & Target & Area Increased \\
    \midrule
    \cite{yao2021programmable} & 4608  & 1152  & 0     & 4.10\% & AES   & 175\% / 69.0\% \\
    \cite{zhang2022darpt} & 321   & 258   & 0     & 0.70\% & AES   & 12.1\% / 15.3\% \\
    \cite{dubey2020bomanet} & 8000  & 6499  & 163   & 8.63\% & BNN   & 430\% / 580\% \\
    \cite{dubey2022guarding} & 9818  & 7709  & 163   & 10.43\% & BNN   & 540\% / 680\% \\
    \NAME  & 8251  & 170   & 0     & 1.28\% & DNN   & \textbf{10.1\%} / \textbf{0.1\%} \\
    \bottomrule
    \end{tabular}}%
  \label{util}%
\end{table}%

\subsection{Defense Effectiveness}
\label{sec:defense-effectiveness}
\begin{figure}[t]
	\centering
        \includegraphics[scale=0.35,trim=0 0 0 0]{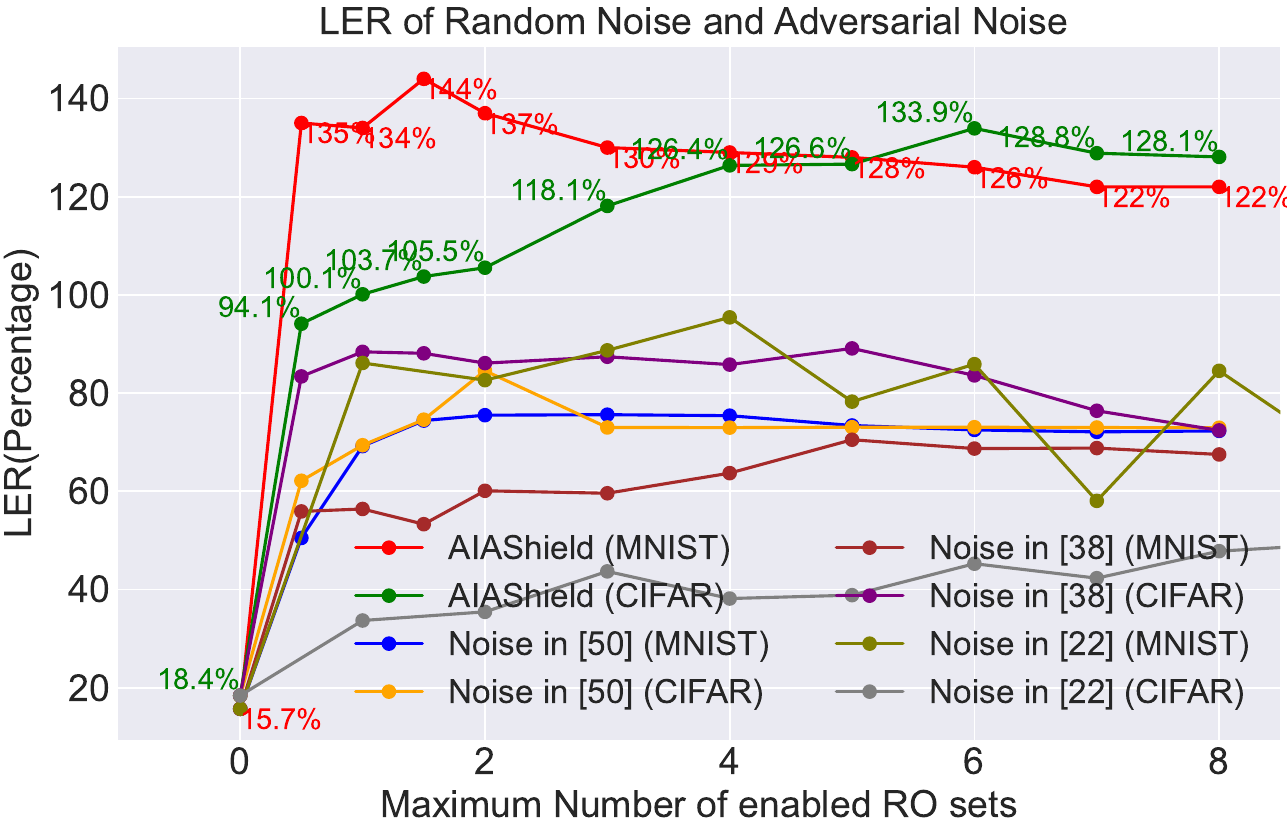}
	\caption{LERs for random noise and our adversarial noise implementations over two datasets}
    \vspace{-15pt}
	\label{LER1}
\end{figure}

\noindent\textbf{Model Similarity Reduction.} We begin by evaluating the effectiveness of our Model Similarity Reduction defense strategy. In this approach, we calculate the adversarial noise in \NAME with varying RO enable counts through our Noise Generation Module. Figure \ref{LER1} compares the LER of \NAME and baseline methods under different maximum numbers of enabled RO sets. We observe that the LERs of our delicate adversarial noise are significantly higher than that of the random noise in other methods over both CIFAR-10 and MNIST datasets for all enable counts of ROs. 

It reveals that other kind of noise is notably less effective than adversarial noise in \NAME.

\noindent\textbf{Model Utility Reduction.}
We now shift our attention to the Model Utility Reduction defense strategy. 
We use our adapted NAS algorithm to identify the worst-performing architecture, apply the adapted universal PGD algorithm to calculate the targeted noise, and then use our noise generator to inject the noise into the model execution.  
The maximum enabled number of RO sets is set to 32 for all defense methods. 
The evaluation results are summarized in Tables \ref{target1} and \ref{target2}, for the MNIST and CIFAR10 datasets, respectively. Acc (Label) refers to the average accuracy of the original victim models. Search Acc (Target Model) refers to the accuracy of the identified target model from our adapted NAS, which is the goal to approach with the noise. The details of these target models on both datasets are reported in Table \ref{targetmodel}, where ``conv'' and ``fc'' refer to the convolution layer and fully-connected layer, respectively. Extract Acc (Target Noise) refers to the actual accuracy of attacker's extracted model with the identified targeted adversarial noise implemented on FPGA. For fair comparisons, all models are trained with the same hyperparameters with 15 epochs and batch size of 512. 

\begin{table}[htbp]
  \centering
  \caption{LER and accuracy for the Model Utility Reduction defense on the MNIST dataset}
  \resizebox{\linewidth}{!}{
    \begin{tabular}{lrrrrr}
    \toprule
    Defense & \multicolumn{1}{l}{LER to } & \multicolumn{1}{l}{LER to } & \multicolumn{1}{l}{Acc } & \multicolumn{1}{l}{Search Acc } & \multicolumn{1}{l}{Extract Acc }\\
       \multicolumn{1}{l}{Methods}    & \multicolumn{1}{l}{label} & \multicolumn{1}{l}{target} & \multicolumn{1}{l}{(Label)} & \multicolumn{1}{l}{(Target Model)} & \multicolumn{1}{l}{(Target Noise)}\\
    \midrule
    \cite{zhao2018fpga} & 94.6\%  & 356.3\%  & 95.4\% & - & 96.1\%\\
    \cite{pothukuchi2021maya} & 73.3\%  & 153.7\%  & 95.4\% & - & 96.2\%\\
    \cite{krautter2019active} & 98.3\%  & 257.4\%  & 95.4\% & - & 97.3\%\\
    \NAME & 98.1\%  & 44.3\%  & 95.4\% & 92.5\% & \textbf{93.8\%}\\

    \bottomrule
    \end{tabular}}%
  \label{target1}%
\end{table}%

\begin{table}[htbp]
  \centering
  \caption{LER and accuracy for the Model Utility Reduction defense on the CIFAR dataset}
  \resizebox{\linewidth}{!}{
    \begin{tabular}{lrrrrr}
    \toprule
    Defense & \multicolumn{1}{l}{LER to } & \multicolumn{1}{l}{LER to } & \multicolumn{1}{l}{Acc } & \multicolumn{1}{l}{Search Acc } & \multicolumn{1}{l}{Extract Acc }\\
       \multicolumn{1}{l}{Methods}    & \multicolumn{1}{l}{label} & \multicolumn{1}{l}{target} & \multicolumn{1}{l}{(Label)} & \multicolumn{1}{l}{(Target Model)} & \multicolumn{1}{l}{(Target Noise)}\\
    \midrule
    \cite{zhao2018fpga} & 37.1\%  & 263.7\%  & 60.3\% & - & 51.8\%\\
    \cite{pothukuchi2021maya} & 74.9\%  & 140.0\%  & 60.3\% & - & 63.0\%\\
    \cite{krautter2019active} & 74.3\%  & 124.8\%  & 60.3\% & - & 51.5\%\\
    \NAME & 75.3\%  & 69.1\%  & 60.3\% & 40.0\% & \textbf{42.9\%}\\

    \bottomrule
    \end{tabular}}%
  \label{target2}%
\end{table}%

\begin{table}[htbp]
  \centering
  \caption{The details of the searched (target) models with the worst performance}
    \begin{tabular}{llr}
    \toprule
    Dataset & Target Model Architecture & \multicolumn{1}{l}{Search Acc} \\
    \midrule
    MNIST & conv(kernel=4*4,output fmap=30)-fc-fc & 92.5\% \\
    CIFAR & conv(kernel=2*2,output fmap=10)-fc & 40.0\% \\
    \bottomrule
    \end{tabular}
  \label{targetmodel}%
\end{table}%

As observed in Tables \ref{target1} and \ref{target2}, the extracted accuracy of the attacker's model is very close to the searched accuracy for both MNIST and CIFAR-10, which indicates that \NAME can generate and implement the noise as precisely as expected. Besides, the extracted accuracy is much lower than that of the victim model, proving the model utility is reduced. We also have two interesting observations. First, for other defense baselines, their extracted accuracy can surpass the actual accuracy of the victim model in MNIST. This is because these methods try to reduce the similarity between the victim and extracted models, while overlooking the potential for the extracted models to exhibit superior accuracy, rendering the attacks still effective. Second, for our \NAME, the model accuracy reduction is more prominent in the CIFAR-10 dataset, where the attacker's model accuracy is almost 20\% below the victim model. In the MNIST dataset, the extracted accuracy is slightly lower than the victim model ($\approx 2\%$), which is attributed to the task's simplicity, as even the searched model can still yield acceptable accuracy. In conclusion, we believe \NAME will cause satisfactory accuracy degradation in practice for handling complex tasks.

\subsection{Defense Transferability}
\label{transferabilitysection}
\NAME operates in a black-box setting by generating the adversarial noise from a surrogate model instead of the actual attack one. We proceed to assess its transferability for two defense strategies. This involves evaluating their efficacy against different, potentially unknown, attacker models. We construct five models with varying structures in Table \ref{trans}. Model 0 is the surrogate model used by the defender to generate the adversarial noise, while models 1-4 used by the attacker for side-channel analysis. We consider different noise intensity configurations, which are manipulated via varying the enable numbers of RO sets

\begin{table}[htbp]
  \centering
  \caption{Details of the defender's surrogate model and attacker's actual models}
    \begin{tabular}{lll}
    \toprule
    Model id & Structure & Details  \\
    \midrule
    model 0 & CNN-RNN-CTC & 3 CNN layers, 2 RNN layers\\
    model 1 & CNN-RNN-CTC & 1 CNN layer, 2 RNN layers\\
    model 2 & CNN-RNN-CTC & 2 CNN layer, 2 RNN layers\\
    model 3 & CNN-RNN-CTC & 5 CNN layer, 2 RNN layers\\
    model 4 & Transformer & 1 encoder and 1 decoder layer\\
    \bottomrule
    \end{tabular}%
  \label{trans}%
\end{table}%

\begin{figure*}
  \centering
		\subfloat[Results for MNIST dataset\label{trans_result:a}]{
			\includegraphics[scale=0.28]{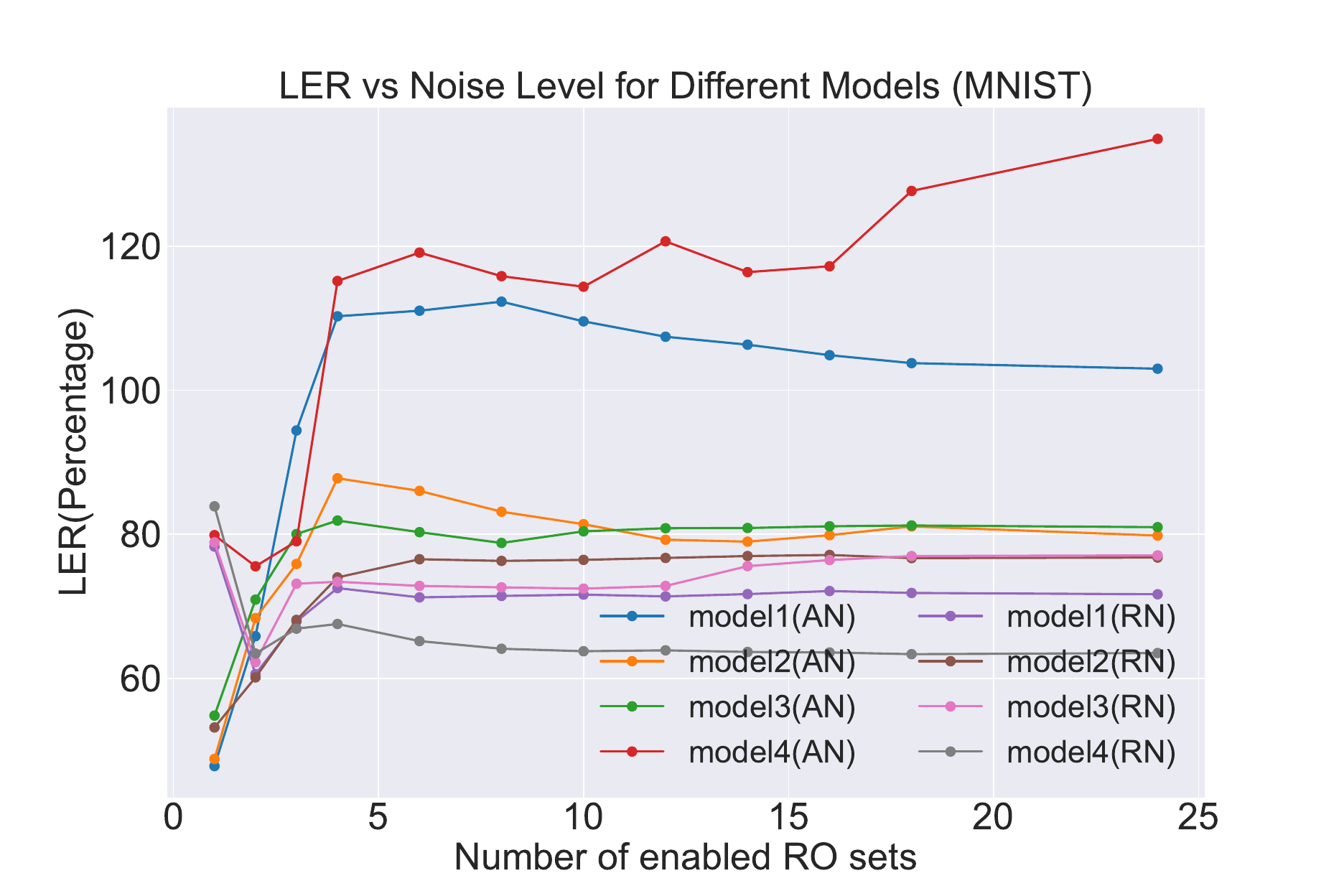}}
		\subfloat[Results for CIFAR dataset\label{trans_result:b}]{
			\includegraphics[scale=0.32]{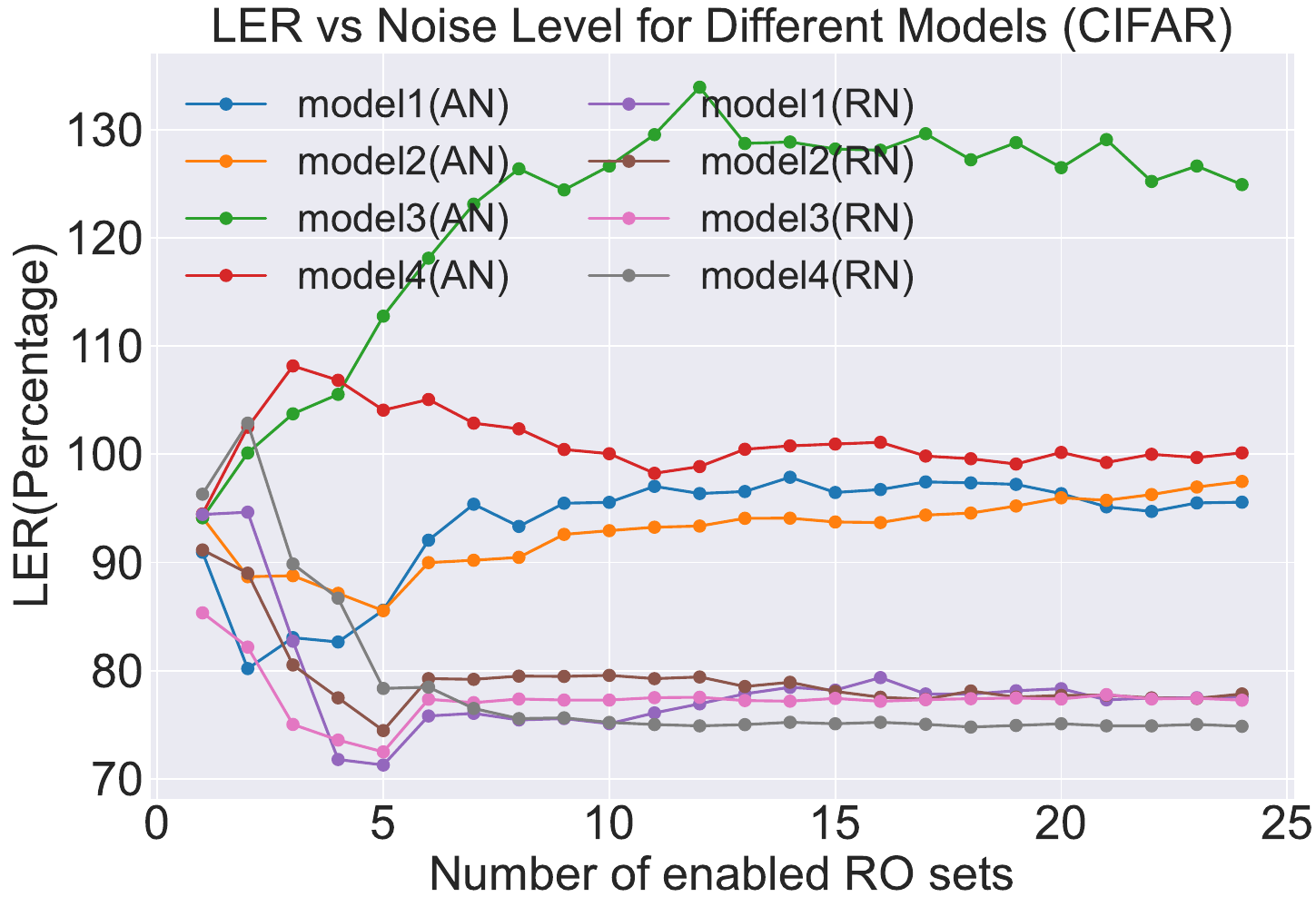}}
\caption{LERs for different attack models. ``AN'': adversarial noise from \NAME; ``RN'': random noise}
\vspace{-10pt}
\label{trans_result}
\end{figure*}

\noindent\textbf{Model Similarity Reduction.}
To assess the transferability of the Model Similarity Reduction defense, 

we measure the LERs between the extracted and victim model in Figure \ref{trans_result}.
We observe that, across both datasets, \NAME induces a significantly higher error rate in the extracted models compared to the utilization of random noise across the four distinct models. This observation underscores the robustness and efficacy of \NAME against distinct attack models.

\noindent\textbf{Model Utility Reduction.}
We proceed to test the transferability of the Model Utility Reduction defense strategy.

The evaluation results for MNIST and CIFAR10 are shown in Tables \ref{trans_result2} and \ref{trans_result3}, respectively. For fair comparisons, all models are trained with the same hyperparameters with 15 epochs and batch size of 512. These tables clearly show that the extracted accuracy of the attacker's model is lower than the victim's model accuracy for all the four cases. In model 2, the defense result is even better than using the surrogate model (model 0) for extraction. This confirms the transferability and effectiveness of \NAME against different attack models. 

\begin{table}[htbp]
  \centering
  \caption{Transferability results of the Model Utility Reduction defense on MNIST}
    \begin{tabular}{lrrr}
    \toprule
    Model id & \multicolumn{1}{l}{LER} & \multicolumn{1}{l}{Acc} & \multicolumn{1}{l}{Extract Acc } \\
          & \multicolumn{1}{l}{(to target)} & \multicolumn{1}{l}{(ground truth)} & \multicolumn{1}{l}{(Target Noise)} \\
    \midrule
    model 1 & 73.5\% & 95.4\% & 94.7\% \\
    model 2 & 11.7\% & 95.4\% & 92.4\% \\
    model 3 & 80.9\% & 95.4\% & 94.6\% \\
    model 4 & 62.3\% & 95.4\% & 94.6\% \\
    \bottomrule
    \end{tabular}%
  \label{trans_result2}%
\end{table}%

\begin{table}[htbp]
  \centering
  \caption{Transferability results of the Model Utility Reduction defense on CIFAR-10}
    \begin{tabular}{lrrr}
    \toprule
    Model id & \multicolumn{1}{l}{LER} & \multicolumn{1}{l}{Acc} & \multicolumn{1}{l}{Extract Acc } \\
          & \multicolumn{1}{l}{(to target)} & \multicolumn{1}{l}{(ground truth)} & \multicolumn{1}{l}{(Target Noise)} \\
    \midrule
    model 1 & 91.6\% & 60.3\% & 49.8\% \\
    model 2 & 78.0\% & 60.3\% & 45.1\% \\
    model 3 & 81.8\% & 60.3\% & 48.3\% \\
    model 4 & 97.5\% & 60.3\% & 49.9\% \\
    \bottomrule
    \end{tabular}%
  \label{trans_result3}%
\end{table}%

\subsection{Defense Robustness}
\begin{figure}[t]
	\centering
        \includegraphics[scale=0.25,trim=0 0 0 0]{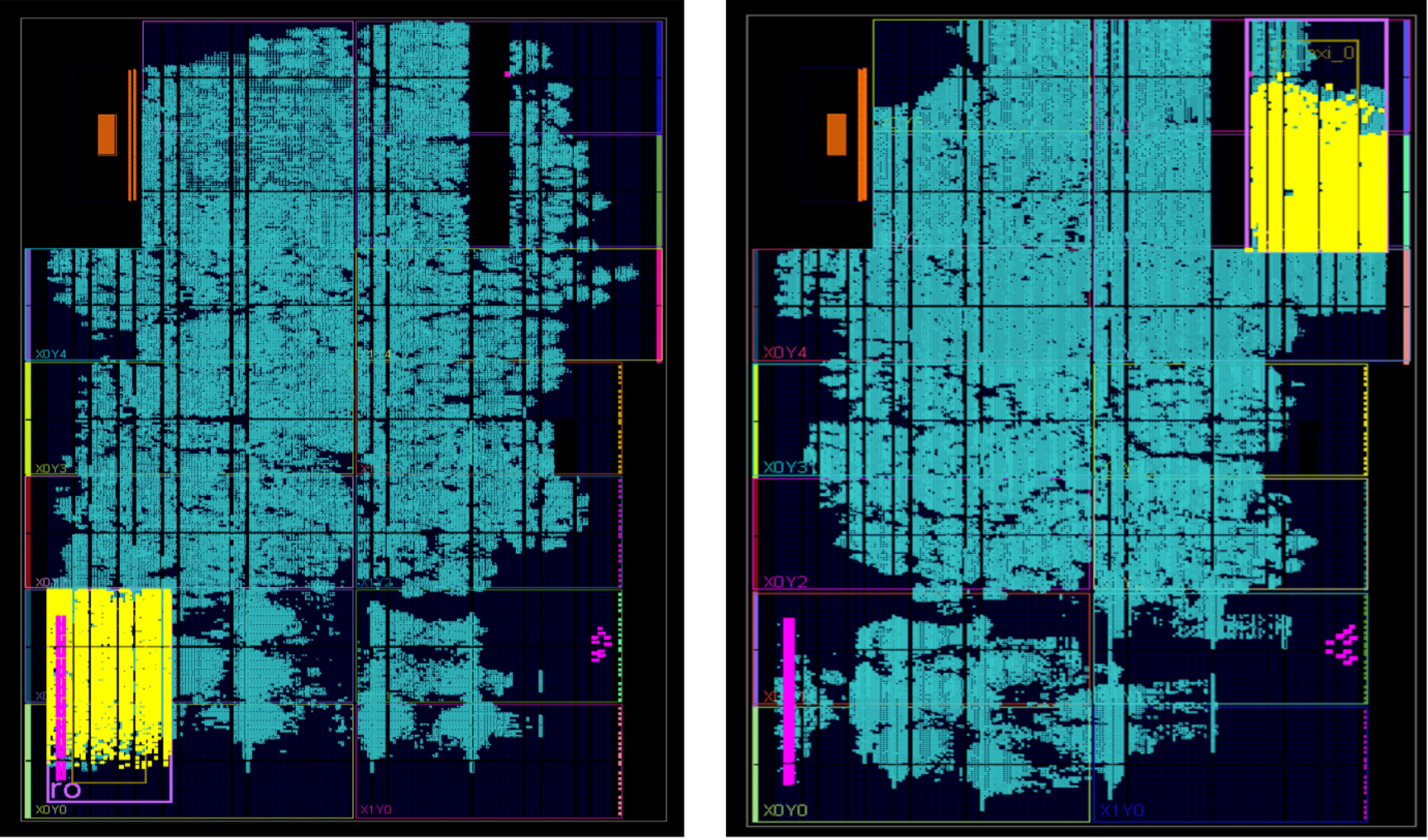}
	\caption{Floorplans of the original (left) and far (right) setups. The Noise Generation Module is shown in yellow, the TDC is shown in purple, and NVDLA is shown in blue. }
	\label{floorplan}
	
\end{figure}

We explore the robustness of \NAME against the locations of the Noise Generation Module. Figure \ref{floorplan} illustrates the floorplans of the original configuration (left) where the attacker's TDC sensor and defender's Noise Generation Module are near, and the far setup (right) where the Noise Generation Module is positioned away from the TDC sensor. We conduct experiments using both datasets to assess the influence of this variation. The results for the two defense strategies are shown in Tables \ref{far1} and \ref{far2}, respectively.

\begin{table}[htbp]
  \centering
  \caption{LERs for the Model Similarity Reduction defense (far setup) over two datasets}
  \resizebox{\linewidth}{!}{
    \begin{tabular}{lrrrrrrrr}
    \toprule
    Max enabled & 1        & 4     & 5    & 7    & 10    & 12   & 14\\
    RO sets &             &       &       &       &       &       & \\
    \midrule
    MNIST & 80.9    & 98.6  & 104.6  & 109.0  & 107.3  & 105.8  & 106.9\\
    CIFAR & 76.0   & 85.4  & 85.1  & 97.0  & 99.0  & 100.6  & 101.5\\

    \bottomrule
    \end{tabular}}%
  \label{far1}%
\end{table}%

\begin{table}[htbp]
  \centering
  \caption{LER and accuracy for the Model Utility Reduction defense (far setup) over two datasets}
  \resizebox{\linewidth}{!}{
    \begin{tabular}{lrrrrr}
    \toprule
     \multicolumn{1}{l}{Dataset} &\multicolumn{1}{l}{LER to } & \multicolumn{1}{l}{LER to } & \multicolumn{1}{l}{Acc } & \multicolumn{1}{l}{Exp. Acc } & \multicolumn{1}{l}{Impl. Acc } \\
        \multicolumn{1}{l}{} &\multicolumn{1}{l}{label} & \multicolumn{1}{l}{target} & \multicolumn{1}{l}{(label)} & \multicolumn{1}{l}{(Target Model)} & \multicolumn{1}{l}{(Target Noise)} \\
    \midrule
    MNIST & 67.2\%  & 15.3\%  & 95.4\% & 92.5\% & 93.2\% \\
    CIFAR & 73.4\%  & 70.0\%  & 60.3\% & 40.0\% & 42.0\% \\

    \bottomrule
    \end{tabular}}%
  \label{far2}%
\end{table}%
These two tables demonstrate that the efficacy of \NAME remains robust with different locations of the Noise Generation Module on the FPGA. It can effectively obfuscate the attacker's observations regardless of the attacker's TDC sensors. This resilience highlights the robust nature of our approach and its potential to counteract model extraction attacks across diverse real-world scenarios.

\section{Related Works}
\label{sec:relatedworks}
\subsection{Protecting AI Accelerators from Power SCAs}
Although lots of works have focused on the power-based SCAs, little efforts are made towards the defenses against ML-based SCAs, which are more powerful and harder to mitigate. Dubey et al. \cite{dubey2020bomanet} designed masked circuits to the component in the neural network such as adder, activation function, multiplexer, and output layer. Each input is split into multiple shares in the mask circuits. A revised masking scheme is further proposed in \cite{dubey2022guarding}, which adds shuffling and the masked design for a baseline Kogge Stone adder (KSA) architecture. However, masking can only provide weight protection to lower-order attacks, while failing to resist attacks that recover model architectures, or attacks with higher-order attacker abilities. Besides, it only works for BNN due to the non-linearity of activation function in DNN. Maji et al. \cite{maji2022threshold} introduced a masked Integrated Circuit (IC) solution, incorporating protection against power and electromagnetic SCAs. This solution involves a threshold implementation (TI) with a 64\% area overhead and 5.5× energy overhead. Trivium stream cipher is employed to generate the necessary random numbers for the TI. Unfortunately, this approach remains susceptible to the challenges outlined earlier. Hashemi et al. \cite{hashemi2022hwgn2} proposed an NN implementation named HWGN2, leveraging garbled circuits. This approach relies on private function evaluation (PFE) and secure function evaluation (SFE) techniques to enhance side-channel resistance. Dubey et al. \cite{dubey2023hardware} developed a secure RISC-V-based co-processor that can execute a neural network implemented in C/C++. The co-processor uses masking to execute various neural network operations like weighted summations, activation functions, and output layer computation in a side-channel secure fashion.

\subsection{Protecting Cryptographic Accelerators from Power SCAs}
Other than DNN accelerators, researchers also propose countermeasures for Cryptographic Accelerators. Zhang et al. \cite{zhang2022darpt} used TDC and pseudo-random number generator (PRNG) to produce random clock jitters to hide sensitive information in the power trace. The experiments on Advanced Encryption Standard (AES) demonstrate that up to 800k traces (100 times) are required to recover the key with correlation power analysis (CPA). Ahmadi et al, \cite{morid2023shield} proposed a defense scheme which exploits ROs to hide the sensitive information in the time domain. It utilizes an offline pre-processing stage to set the number of RO counters, the placement of the Power Monitor, and the sampling frequency of the Power Monitor before being implemented on board. Furthermore, Krautter et al. \cite{krautter2019active} introduced an RO array named ``active fence'' controlled by sensors to minimize the voltage fluctuations. Their evaluations reveal that the effectiveness of initiating CPA against AES necessitates a substantial number of traces, around 300k (167 times). However, this approach incurs notable overhead, with a doubling of the FPGA area and a 50\% increase in power consumption. Pothukuchi et al. \cite{pothukuchi2021maya} addressed power side-channel attacks on CPUs by reshaping the power dissipation using formal control techniques, employing tasks like the ``balloon task'' and manipulating the idle activity. Gaussian sinusoid noise is applied to obscure the power information, demonstrating a unique approach to defense in this context.

\section{Discussion and Future Works}
\label{sec:discussion}

In this section, we discuss how to extend \NAME to other settings and scenarios. 

\noindent\textbf{Extension to Other Side Channels.}
While originally designed to counteract power side channels, \NAME exhibits versatility in defeating other types of side channels. For instance, Gupta et al. \cite{cryptoeprint:2023/368} exploited electromagnetic (EM) traces to extract model architectures from NVDLA. As the employed ML models are also susceptible to adversarial attacks, we can similarly craft the adversarial noise and inject it into attacker's EM observations. To adapt \NAME to a new side channel, the primary adjustment involves customizing the Monitor and Noise Generation Module to generate and incorporate the noise specific to the characteristics of the side channel.

\noindent\textbf{Extension to Other Target Applications.}
\NAME possesses the capability to safeguard other applications against ML-based side-channel attacks. For instance, Kubota et al. \cite{KUBOTA2021103383} proposed to use a CNN model to analyze the power trace from the execution of AES and recover the key bytes. We can use \NAME to generate and inject the adversarial noise into the power traces, and prevent the key leakage.

\noindent\textbf{Extension to Other Hardware Platforms.}
In addition to hardware accelerators, the design of \NAME can also be extended to other platforms, with the adjustment of Monitor and Noise Generation Module. For instance, Luo et al. \cite{7357115} presented a power SCA on Nvidia Tesla GPU. We can apply the algorithms in \NAME to generate targeted or untargted noise. Then we can adopt a Keysight oscilloscope as the Power Monitor to assess the voltage drop on the resistor, and use parallel computation of other programs as the Noise Generator Module to inject the delicate noise. 

\noindent\textbf{Extension to Larger-scale Models.}
As shown in Section \ref{sec:defense-effectiveness}, the Model Utility Reduction defense is more effective for larger models. So we believe \NAME is general to protect more complex AI models. In these sophisticated tasks, the attacker may necessitate gathering more side-channel data, achievable by elevating the frequency of the power monitor. The defender, in response, can also elevate the frequency of the Noise Generation Module by increasing the frequency of the AXI bus in our design to inject a greater amount of noise into obfuscate the attacker's observations. 

\section{Conclusion}
\label{sec:conclusion}
In this paper, we present \NAME, a novel hardware-based defense approach to protect AI accelerators from power SCAs. Our key idea is to innovatively apply the adversarial attack techniques in ML security to generate delicate noise, which can effectively obfuscate the attacker's side-channel observation, while incurring minimizing impact on the FPGA board and model execution. We introduce two defense goals: in \textit{Model Similarity Reduction} defense, we adopt the untargeted adversarial attack algorithm to reduce the resemblance between the attacker's recovered model and victim's model in terms of extraction error rate; in \textit{Model Utility Reduction} defense, we leverage the targeted adversarial attack algorithm with NAS to entice the attacker to extract a model with poor performance. We further design a hardware module with a software calibration scheme to generate fine-grained noise for side-channel obfuscation. Our comprehensive evaluation demonstrates that \NAME effectively withstands attacks with minimal overhead, and exhibits remarkable transferability and robustness across a variety of attack models and sensor locations employed by the attacker.

\bibliographystyle{plain}
\bibliography{mybib}

\end{document}